\documentclass[letterpaper,twocolumn,twoside]{IEEEtran}
\usepackage{amssymb, cite, epsfig, amsmath}

\def\beq{\begin{equation}}
\def\eeq{\end{equation}}
\def\beqa{\begin{eqnarray}}
\def\eeqa{\end{eqnarray}}
\def\beqan{\begin{eqnarray*}}
\def\eeqan{\end{eqnarray*}}
\def\R{{\mathbb{R}}}
\def\argmin{\mathop{\mathrm{arg\,min}}}
\def\dim{\mathop{\mathrm{dim}}}
\def\range{\mathop{\mathrm{range}}}
\def\limn{\lim_{n \rightarrow \infty}}
\def\liminfn{\liminf_{n \rightarrow \infty}}
\def\limsupn{\limsup_{n \rightarrow \infty}}

\newtheorem{definition}{Definition}
\newtheorem{theorem}{Theorem}
\newtheorem{lemma}{Lemma}

\def\Ihat{\ensuremath{\hat{I}}}
\def\IhatSUD{\ensuremath{\hat{I}_{\rm T}}}
\def\IhatSOMP{\ensuremath{\hat{I}_{\rm S}}}
\def\Itrue{\ensuremath{I_{\rm true}}}
\def\Perr{\ensuremath{p_{\rm err}}}
\def\PMD{\ensuremath{p_{\rm MD}}}
\def\PFA{\ensuremath{p_{\rm FA}}}
\def\gammaOpt{\ensuremath{\gamma_{\rm opt}}}
\def\SNR{\mbox{\small \sffamily SNR}}
\def\MAR{\mbox{\small \sffamily MAR}}
\def\tableSNR{\mbox{\tiny \sffamily SNR}}
\def\tableMAR{\mbox{\tiny \sffamily MAR}}
\def\captionSNR{\mbox{\scriptsize \sffamily SNR}}
\def\captionMAR{\mbox{\scriptsize \sffamily MAR}}
\def\arr{\rightarrow}
\def\Exp{\mathbf{E}}
\def\sigmahat{\ensuremath{\widehat{\sigma}}}
\def\SNRmin{\mbox{\small \sffamily SNR}_{\rm min}}
\def\dB{{\rm dB}}
\def\BetaDist{\mbox{Beta}}
\def\Ptrue{\ensuremath{\mathbf{P}_{\rm true}}}
\def\rhotrue{\ensuremath{\rho_{\rm true}}}
\def\mmin{m_{\rm min}}
\def\smin{s_{\rm min}}
\def\zmax{s_{\rm max}}

\newcommand{\abf}{\mathbf{a}}
\newcommand{\bbf}{\mathbf{b}}
\newcommand{\dbf}{\mathbf{d}}
\newcommand{\ebf}{\mathbf{e}}
\newcommand{\sbf}{\mathbf{s}}
\newcommand{\vbf}{\mathbf{v}}
\newcommand{\xbf}{\mathbf{x}}
\newcommand{\xbfhat}{\widehat{\mathbf{x}}}
\newcommand{\ybf}{\mathbf{y}}
\newcommand{\Abf}{\mathbf{A}}
\newcommand{\Pbf}{\mathbf{P}}
\newcommand{\etal}{\emph{et al.}}

\begin{document}

\title{Ranked Sparse Signal Support Detection}
\author{Alyson K. Fletcher,
        Sundeep Rangan,
        and Vivek K Goyal
}

\markboth{Ranked Sparse Signal Support Detection}
        {Fletcher, Rangan and Goyal}

\maketitle

\begin{abstract}
This paper considers the problem of detecting the support
(sparsity pattern) of a sparse vector from random noisy measurements.
Conditional power of a component of the sparse vector is defined as the 
energy conditioned on the component being nonzero.
Analysis of a simplified version of orthogonal matching pursuit
(OMP) called sequential OMP (SequOMP) demonstrates the importance
of knowledge of the rankings of conditional powers.
When the simple SequOMP algorithm is applied to components in
nonincreasing order of conditional power, the detrimental effect
of dynamic range on thresholding performance is eliminated.
Furthermore, under the most favorable conditional powers,
the performance of SequOMP approaches maximum likelihood performance
at high signal-to-noise ratio.
\end{abstract}


\section{Introduction}
Sets of signals that are sparse or approximately sparse with respect
to some basis are ubiquitous because signal modeling often has the
implicit goal of finding such bases.
Using a sparsifying basis,
a simple abstraction that applies in many settings is for
\beq \label{eq:yax}
    \ybf \ = \ \Abf \xbf + \dbf
\eeq
to be observed,
where $\Abf \in \R^{m \times n}$ is known,
$\xbf \in \R^n$ is the unknown sparse signal of interest, and
$\dbf \in \R^m$ is random noise.
When $m < n$, constraints or prior information about $\xbf$ are
essential to both estimation
(finding vector $\xbfhat(\ybf)$ such that $\| \xbf - \xbfhat \|$ is small)
and detection
(finding index set $\Ihat(\ybf)$ equal to the support of $\xbf$).
The focus of this paper is on the use of magnitude rank
information on $\xbf$---in addition to sparsity---in the support detection
problem.
We show that certain scaling laws relating the problem dimensions and
the noise level are changed dramatically by exploiting the rank information
in a simple sequential detection algorithm.

The simplicity of the observation model (\ref{eq:yax}) belies the
variety of questions that can be posed and the difficulty of precise analysis.
In general, the performance of any algorithm is a complicated function of
$\Abf$, $\xbf$, and the distribution of $\dbf$.
To enable results that show the qualitative behavior in terms of
problem dimensions and a few other parameters,
we assume the entries of $\Abf$ are i.i.d.\ normal
and describe $\xbf$ by its
energy and its smallest-magnitude nonzero entry.

We consider a partially-random signal model
\beq \label{eq:xmodel}
  x_j \ = \ b_j \, s_j,
\qquad
  j = 1,\,2,\,\ldots,\,n,
\eeq
where components of vector $\bbf$
are i.i.d.\ Bernoulli random variables with 
$\Pr(b_j = 1) = 1-\Pr(b_j=0) = \lambda > 0$
and $\sbf$ is a nonrandom parameter vector with all nonzero entries.
The value $s_j^2$ represent the \emph{conditional power} of the component
$x_j$ in the event that $b_j=1$.
We consider the problem where the estimator knows neither $b_j$
nor $s_j$, but may know the \emph{order} or \emph{rank}
of the conditional powers.
In this case, the estimator can, for example, sort the components
of $\sbf$ in an order such that 
\beq \label{eq:ranks}
  |s_1| \ \geq \ |s_2| \ \geq \ \cdots \ \geq \ |s_n| \ > \ 0.
\eeq

The main contribution of this paper is to show that this 
rank information is extremely valuable.  
A stylized application in which the conditional ranks can be known
is random access communication as described in~\cite{FletcherRG:09a}.
Irrespective of this application, we show that
when conditional rank information is available, a very simple
detector, termed \emph{sequential orthogonal matching pursuit} (SequOMP),
can be effective.  The SequOMP algorithm is a one-pass version
of the well-known orthogonal matching pursuit (OMP) algorithm (see references
below).  Similar to several works in sparsity pattern recovery 
\cite{Wainwright:09-ml,Wainwright:09-lasso,FletcherRG:09-IT},
we analyze the performance of SequOMP by estimating a scaling on the 
minimum number of measurements $m$ to asymptotically reliably detect
the sparsity pattern (support) of $\xbf$ in the limit of large
random matrices $\Abf$.
Although the SequOMP algorithm is extremely simple, we show:
\begin{itemize}
\item When the power orders are known and the signal-to-noise ratio (SNR)
is high, the SequOMP algorithm exhibits
a scaling in the minimum number of measurements 
for sparsity pattern recovery that is within a constant factor
of the more sophisticated lasso and OMP algorithms.
In particular, SequOMP exhibits a resistance to 
large dynamic ranges, which is one of the main motivations for using
lasso and OMP\@.
\item When the power profile can be optimized, SequOMP
can achieve measurement scaling for sparsity pattern recovery that
is within a constant factor of optimal ML
detection.  This scaling is better than the best known 
sufficient conditions for lasso and OMP\@.
\end{itemize}
The results are not meant to suggest that SequOMP
is a good algorithm in any sense:  other algorithms such as OMP
can perform dramatically better.  The point is to
concretely and provably demonstrate the value of conditional rank information.

\subsection{Related Work}
Under an i.i.d.\ Gaussian assumption on $\dbf$, maximum likelihood estimation of
$\xbf$ under
a sparsity constraint
is equivalent to finding sparse $\xbfhat$ such that $\|\ybf - \Abf \xbfhat\|_2$
is minimized.  This is called optimal sparse approximation of $\ybf$
using dictionary $\Abf$, and it is NP-hard~\cite{Natarajan:95}.
Several greedy heuristics (matching pursuit~\cite{MallatZ:93} and its
variants with orthogonalization~\cite{ChenBL:89,PatiRK:93,DavisMZ:94}
and iterative refinement~\cite{NeedellT:09,DaiM:09})
and convex relaxations (basis pursuit~\cite{ChenDS:99},
lasso~\cite{Tibshirani:96}, Dantzig selector~\cite{CandesT:07}, and others)
have been developed for sparse approximation,
and under certain conditions on $\Abf$ and $\ybf$
they give optimal or near-optimal
performance~\cite{DonohoET:06,Tropp:04,Tropp:06}.
Results showing that near-optimal estimation of $\xbf$ is obtained with
convex relaxations,
pointwise over compressible $\xbf$ and
with high probability over some random ensemble for $\Abf$,
form the heart of the compressed sensing
literature~\cite{CandesRT:06-IT,Donoho:06,CandesT:06}.
Under a probabilistic model for $\xbf$ and certain additional assumptions,
exact asymptotic performances of several estimators are
known~\cite{RanganFG:0x-IT}.

Our interest is in recovery or detection of the support
(or \emph{sparsity pattern}) of $\xbf$ rather than the estimation of $\xbf$.
In the noiseless case of $\dbf = 0$, optimal estimation of $\xbf$
can yield $\xbfhat = \xbf$ under certain conditions on $\Abf$;
estimation and detection then coincide, and some papers cited above and
notably~\cite{DonohoT:09} contain relevant results.
In the general noisy case, direct analysis of the detection problem
has yielded much sharper results.

A standard formulation is to treat $\sbf$ as a nonrandom parameter vector
and $\bbf$ as either nonrandom with weight $k$ or random with a uniform
distribution over the weight-$k$ vectors.
The minimum probability of detection error is then attained with
maximum likelihood (ML) detection.  Sufficient conditions for the
success of ML detection are due to Wainwright~\cite{Wainwright:09-ml};
necessary conditions based on channel capacity were given by several
authors~\cite{SarvothamBB:06-Allerton,FletcherRG:07b,Reeves:08,AkcakayaT:10},
and conditions more stringent in many regimes and a comparison of results
appears in~\cite{FletcherRG:09-IT}.
Necessary and sufficient conditions for lasso were determined by
Wainwright~\cite{Wainwright:09-lasso}.
Sufficient conditions for orthogonal matching pursuit (OMP) were given by
Tropp and Gilbert~\cite{TroppG:07} and improved by
Fletcher and Rangan~\cite{FletcherR:09}.
Even simpler than OMP is a thresholding algorithm analyzed in
a noiseless setting in~\cite{RauhutSV:08}
and with noise in~\cite{FletcherRG:09-IT}.
These results are summarized in Table~\ref{table:summary},
using terminology defined formally in Section~\ref{sec:chanMod}.

\subsection{Paper Organization}
The remainder of the paper is organized as follows.
The setting is formalized in Section~\ref{sec:chanMod}.
In particular, we define all the key problem parameters.
Common algorithms and previous results on their performances
are then presented in Section~\ref{sec:csAnalysis}.
We will see that there is a potentially-large performance gap
between the simplest thresholding algorithm and the optimal ML detection,
depending on the signal-to-noise ratio (SNR) and the dynamic range of $\xbf$.
Section~\ref{sec:SOMP} presents a new detection algorithm,
sequential orthogonal matching pursuit (SequOMP),
that exploits knowledge of conditional ranks.
Numerical experiments are reported in Section~\ref{sec:sim}.
Conclusions are given in Section~\ref{sec:concl},
and proofs are relegated to the Appendix.

\section{Problem Formulation}
\label{sec:chanMod}

In the observation model $\ybf = \Abf \xbf + \dbf$,
let $\Abf \in \R^{m \times n}$ and
$\dbf \in \R^{n}$ have i.i.d.\ $\mathcal{N}(0,1/m)$ entries.
This is a normalization under which the ratio of conditional total signal energy
to total noise energy
\beq \label{eq:snrDef}
    \SNR(\xbf) \ = \ \frac{\Exp[\|\Abf\xbf\|^2 \mid \xbf]}
                     {\Exp[\|\dbf\|^2]}
\eeq
simplifies to
\beq \label{eq:snrVal}
    \SNR(\xbf) \ = \ \|\xbf\|^2.
\eeq

Let
$$
    \Itrue \ = \ \left\{~ j \in \{1,2,\ldots,n\} ~:~ x_j \neq 0 ~\right\}
$$
denote the \emph{support}
of $\xbf$.
Using signal model (\ref{eq:xmodel}),
$$
    \Itrue \ = \ \left\{~ j \in \{1,2,\ldots,n\} ~:~ b_j = 1 ~\right\}.
$$
The \emph{sparsity level} of $\xbf$ is
$k = |\Itrue|$.

An estimator produces an estimate $\Ihat = \Ihat(\ybf)$
of $\Itrue$ based on the observed noisy vector $\ybf$.
Given an estimator, its probability of error%
\footnote{An alternative to this definition of $\Perr$ could be to allow a
nonzero fraction of detection errors~\cite{AkcakayaT:10,Reeves:08}.}
\beq \label{eq:Perr}
    \Perr \ = \ \Pr\left( \Ihat \neq \Itrue \right)
\eeq
is taken with respect to randomness in $\Abf$, noise vector $\dbf$,
and signal $\xbf$.
Our interest is in relating the scaling of problem parameters with
the success of various algorithms.  For this, we define the following
criterion.
\begin{definition}
Suppose that we are given deterministic sequences $m = m(n)$,
$\lambda = \lambda(n)$, and $\sbf = \sbf(n) \in \R^n$ that vary with $n$.
For a given detection algorithm $\Ihat = \Ihat(\ybf)$,
the probability of error $\Perr$
is some function of $n$.
We say that the detection algorithm achieves
\emph{asymptotic reliable detection} when $\Perr(n) \arr 0$.
\end{definition}

We will see that two key factors influence the ability to detect $\Itrue$.
The first is the total SNR defined above.
The second is what we call the \emph{minimum-to-average ratio}
\beq \label{eq:MAR-def}
    \MAR(\xbf) \ = \ \frac{\min_{j \in \Itrue} |x_j|^2}
                {{\|\xbf\|^2}/{k}} .
\eeq
Since $\Itrue$ has $k$ elements,
$\|\xbf\|^2 / k$ is the average of $\{ |x_j|^2 : j \in \Itrue\}$.
Therefore, $\MAR(\xbf) \in (0,1]$ with the upper limit occurring when all the nonzero
entries of $\xbf$ have the same magnitude.

Finally, we define the \emph{minimum component SNR} to be
\beq \label{eq:snrMinDef}
    \SNRmin(\xbf) \ = \ \frac{\min_{j \in \Itrue} \Exp[\|\abf_jx_j\|^2 \mid \xbf]}
                   {\Exp[\|\dbf\|^2]}
    \ = \ \min_{j \in \Itrue} |x_j|^2,
\eeq
where $\abf_j$ is the $j$th column of $\Abf$
and the second equality follows from the normalization of chosen
for $\Abf$ and $\dbf$.
The quantity $\SNRmin(\xbf)$ has a natural interpretation:
The numerator
is the signal power due to
the smallest nonzero component in $\xbf$,
while the denominator
is the total
noise power. The ratio $\SNRmin(\xbf)$ thus represents the contribution to
the SNR from the smallest nonzero component of
$\xbf$.
Observe that (\ref{eq:snrVal}) and (\ref{eq:MAR-def}) show
\beq \label{eq:snrMarProd}
    \SNRmin(\xbf) \ = \ \min_{j \in \Itrue} |x_j|^2
        \ = \ \frac{1}{k} \, \SNR(\xbf) \cdot \MAR(\xbf).
\eeq

We will be interested in estimators that exploit minimal prior knowledge on $\xbf$:
either only knowledge of sparsity level (through $k$ or $\lambda$)
or also knowledge of the conditional ranks
(through the imposition of (\ref{eq:ranks})).
In particular, full knowledge of $\sbf$ would change the problem considerably
because the finite number of possibilities for $\xbf$ could be exploited.

\section{Common Detection Methods}
\label{sec:csAnalysis}

In this section, we review several asymptotic analyses for detection of
sparse signal support.
These previous results hold \emph{pointwise} over sequences of problems of
increasing dimension $n$, i.e., treating $\xbf$ as an unknown deterministic
quantity.  That makes these results \emph{stronger} than results
that are limited to the model (\ref{eq:xmodel}) where the $b_j$s are i.i.d.\
Bernoulli variables.
To reflect the pointwise validity of these results,
they are stated in terms of deterministic sequences
$\xbf$, $m$, $k$, $\SNR$, $\MAR$, and $\SNRmin$ that depend on dimension $n$
and are arbitrary aside from satisfying $m \rightarrow \infty$ and
the definitions of the previous section.
To simplify the notation, we drop the dependence of $\xbf$, $m$ and $k$ on $n$,
and $\SNR$, $\MAR$ and $\SNRmin$ on $\xbf(n)$.
When the results are tabulated for comparison with each other and with the
results of Section~\ref{sec:SOMP}, we replace $k$ with $\lambda n$;
this specializes the results to the model (\ref{eq:xmodel}).

\subsection{Optimal Detection with No Noise}
\label{sec:noNoise}

To understand the limits of detection, it is useful to first
consider the minimum number of measurements when there is no noise.
Suppose that $k$ is known to the detector.
With no noise, the observed vector is $\ybf = \Abf \xbf$,
which will belong to one of $J = {n \choose k}$ subspaces
spanned by $k$ columns of $\Abf$.  If $m > k$,
then these subspaces will be distinct with probability 1\@.
Thus, an exhaustive search through
the subspaces will reveal which subspace $\ybf$ belongs to and thus
determine the support $\Itrue$.
This shows that with no noise and no computational limits,
the scaling in measurements of
\beq \label{eq:minMeasOptNoNoise}
    m \ > \ k
\eeq
is sufficient for asymptotic reliable detection.

Conversely, if no prior information is known at the detector
other than $\xbf$ being $k$-sparse, then
the condition (\ref{eq:minMeasOptNoNoise}) is also necessary.
If $m \leq k$, then for almost all $\Abf$,
any $k$ columns of $\Abf$ span $\R^m$.  Consequently, any observed
vector $\ybf = \Abf \xbf$ is consistent with any support of weight $k$.
Thus, the support cannot be determined without further
prior information on the signal $\xbf$.

\subsection{ML Detection with Noise}
Now suppose there is noise.
Since $\xbf$ is an unknown deterministic quantity, the probability of
error in detecting the support is minimized by maximum likelihood
(ML) detection.
Since the noise $\dbf$ is Gaussian, the ML detector finds the $k$-dimensional
subspace spanned by $k$ columns of $\Abf$ containing the maximum energy of $\ybf$.

The ML estimator was first analyzed by Wainwright~\cite{Wainwright:09-ml}.
He shows that there exists a constant $C > 0$ such that if
\beqa
    m &\geq& C
        \max\left\{\frac{1}{\MAR \cdot \SNR}k \log(n-k),
         k\log(n/k) \right\} \nonumber \\
      &=&C\max\left\{\frac{1}{\SNRmin}\log(n-k),
         k\log(n/k) \right\}
      \label{eq:minMeasMLSuff}
\eeqa
then ML will asymptotically detect the correct support.
The equivalence of the two expressions in (\ref{eq:minMeasMLSuff}) is due to
(\ref{eq:snrMarProd}).
Also, \cite[Thm.~1]{FletcherRG:09-IT}
(generalized in \cite[Thm.~1]{WangWR:10})
shows that, for any $\delta > 0$, the condition
\beqa
    m &\geq&
        \frac{2(1-\delta)}{\MAR \cdot \SNR} k \log(n-k)+ k
            \nonumber \\
        &=&\frac{2(1-\delta)}{\SNRmin} \log(n-k)+ k,\label{eq:minMeasML}
\eeqa
is necessary.
Observe that when $\SNR \cdot \MAR \arr \infty$, the lower
bound (\ref{eq:minMeasML}) approaches $m \geq k$, matching the noise-free
case (\ref{eq:minMeasOptNoNoise}) as expected.

These necessary and sufficient conditions for ML appear in
Table~\ref{table:summary} with smaller terms and the infinitesimal $\delta$
omitted for simplicity.

\newlength{\tableVertA}
\setlength{\tableVertA}{1.5mm}
\newlength{\tableVertB}
\setlength{\tableVertB}{1mm}
\newlength{\tableVertC}
\setlength{\tableVertC}{-1.5mm}
\begin{table*}
 \begin{center}
  \begin{tabular}{|c||c|c|}
    \hline
           & finite $\captionSNR \cdot \captionMAR$ &
           $\captionSNR \cdot \captionMAR \rightarrow \infty$ \\ \hline \hline

& & \\[\tableVertC]
Necessary for ML
           & $m >   \frac{2}
                      {\tableMAR \cdot \tableSNR}  k \log(n-k)$
           & $m > k$ \\[\tableVertA]
           & Fletcher \etal \cite[Thm.~1]{FletcherRG:09-IT}
           & (elementary) \\[\tableVertB]
\hline

& & \\[\tableVertC]
Sufficient for ML
           & $m >   \frac{C}
                      {\tableMAR \cdot \tableSNR}  k \log(n-k)$
           & $m > k$ \\[\tableVertA]
           & Wainwright \cite{Wainwright:09-ml}
           & (elementary) \\[\tableVertB]
\hline

& & \\[\tableVertC]
Sufficient for SequOMP
           & $m > \frac{8}{\log(1+\tableSNR)}k \log(n-k)$
           & $m > 9k$ \\[\tableVertA]
with best power profile
           & {\bf From Theorem~\ref{thm:minMeasSOMP}
             (Section~\ref{sec:powShaping})}
           & {\bf From Theorem~\ref{thm:minMeasSOMP}
             (Section~\ref{sec:snrSaturation})} \\[\tableVertB]
\hline

& & \\[\tableVertC]
Sufficient for SequOMP
           & $m > \frac{8(1+\tableSNR\cdot \tableMAR)}{\tableSNR\cdot \tableMAR}k \log(n-k)$
           & $m > 8k\log(n-k)$ \\[\tableVertA]
with known conditional ranks
           & {\bf From Theorem~\ref{thm:minMeasSOMP}
             (Section~\ref{sec:powShaping})}
           & {\bf From Theorem~\ref{thm:minMeasSOMP}
             (Section~\ref{sec:snrSaturation})} \\[\tableVertB]
\hline

& & \\[\tableVertC]
Necessary and
           & complicated; see~\cite{Wainwright:09-lasso}
           & $m > 2 k \log(n-k)$ \\[\tableVertA]
sufficient for lasso
           &
           & Wainwright~\cite{Wainwright:09-lasso} \\[\tableVertB]
\hline

& & \\[\tableVertC]
Sufficient for
           & unknown     & $m > 2 k \log(n-k)$ \\[\tableVertA]
OMP
           &       & Fletcher and Rangan~\cite{FletcherR:09} \\[\tableVertB]
\hline

& & \\[\tableVertC]
Sufficient for
           & $m > \frac{8(1+\tableSNR)}
                       {\tableMAR \cdot \tableSNR} k \log(n-k)$
           & $m > \frac{8}{\tableMAR} k \log(n-k)$ \\[\tableVertA]
thresholding (\ref{eq:IhatSUD})
           & Fletcher \etal \cite[Thm.~2]{FletcherRG:09-IT} & \\[\tableVertB]
\hline

  \end{tabular}
 \end{center}
 \caption{Summary of results on measurement scalings for
   asymptotic reliable detection for various detection algorithms. \newline
   Only leading terms are shown.
   See body for definitions and additional technical limitations.}
 \label{table:summary}
\end{table*}

\subsection{Thresholding} \label{sect:sud}
The simplest method to detect the support is to use a thresholding rule
of the form
\beq \label{eq:IhatSUD}
    \IhatSUD \ = \ \left\{~ j \in \{1,2,\ldots,n\} ~:~ \rho(j) > \mu ~\right\},
\eeq
where $\mu > 0$ is a threshold parameter and $\rho(j)$ is the
correlation coefficient:
$$
    \rho(j) = \frac{|\abf_j'\ybf|^2}{\|\abf_j\|^2\|\ybf\|^2},
\qquad j = 1,\,2,\,\ldots,\,n.
$$
Thresholding has been analyzed
in~\cite{DuarteSBWB:05,RauhutSV:08,FletcherRG:09-IT}.
In particular, \cite[Thm.~2]{FletcherRG:09-IT} is the following:
Suppose
\beqa
    m &>& \frac{2(1 + \delta)(1+\SNR)k \, L(k,n)}{\SNR \cdot \MAR} \nonumber\\
    &=& \frac{2(1 + \delta)(1+\SNR)L(k,n)}{\SNRmin} \label{eq:minMeasSUDFull}
\eeqa
where $\delta > 0$ and
\beq \label{eq:Ldef}
    L(k,n) \ = \ \left[\sqrt{\log(n-k)} + \sqrt{\log(k)}\right]^2.
\eeq
Then there exists a sequence of detection thresholds $\mu = \mu(n)$
such that $\IhatSUD$ achieves asymptotic reliable detection
of the support.
As before, the equivalence of the two expressions in (\ref{eq:minMeasSUDFull}) is due to
(\ref{eq:snrMarProd}).

Comparing the sufficient condition (\ref{eq:minMeasSUDFull})
for thresholding with the \emph{necessary} condition (\ref{eq:minMeasML}),
we see two distinct problems with thresholding:
\begin{itemize}
\item \emph{Constant offset:}  The scaling (\ref{eq:minMeasSUDFull})
for thresholding shows a factor $L(k,n)$ instead of
$\log(n-k)$ in (\ref{eq:minMeasML}).
It is easily verified that, for $k/n \in (0,1/2)$,
\beq \label{eq:lnlamBnd}
    \log(n-k) \ < \ L(k,n) \ < \ 4\log(n-k),
\eeq
so this difference in factors alone could require that thresholding use up to
4 times more measurements than ML for asymptotic reliable detection.

Combining the inequality (\ref{eq:lnlamBnd}) with (\ref{eq:minMeasSUDFull}),
we see that the more stringent, but simpler, condition
\beq \label{eq:minMeasSUD}
    m \ > \ \frac{8(1 + \delta)(1+\SNR)}{\SNR \cdot \MAR} k\log(n-k)
\eeq
is also sufficient for asymptotic reliable detection with thresholding.
This simpler condition is shown in Table~\ref{table:summary},
where we have omitted the infinitesimal $\delta$ quantity to simplify the
table entry.

\item \emph{SNR saturation:}  In addition to the $L(k,n)/\log(n-k)$
offset, thresholding also requires a factor of
$1+\SNR$ more measurements than ML\@.
This $1+\SNR$ factor has a natural interpretation as
\emph{intrinsic interference}:
When detecting any one component of the vector $\xbf$,
thresholding sees the energy from the
other $n-1$ components of the signal as interference.
This interference is distinct from the additive noise $\dbf$,
and it increases the effective noise by a factor of $1+\SNR$.

The intrinsic interference results in a large performance gap at high SNRs.
In particular, as $\SNR \arr \infty$, (\ref{eq:minMeasSUDFull}) reduces to
\beq \label{eq:minMeasSUDlim}
    m \ > \ \frac{2(1 + \delta)k \, L(k,n)}{\MAR}.
\eeq
In contrast, ML may be able to succeed with a scaling $m = O(k)$ for high SNRs.
\end{itemize}

\subsection{Lasso and OMP Detection}

While ML has clear advantages over thresholding,
it is not computationally tractable for large problems.
One practical method is \emph{lasso}~\cite{Tibshirani:96},
also called basis pursuit denoising~\cite{ChenDS:99}.
The lasso estimate of $\xbf$ is obtained by solving the convex optimization
$$
    \xbfhat \ = \ \argmin_{\xbf} \left( \|\ybf - \Abf\xbf\|_2^2 + \mu\|\xbf\|_1 \right),
$$
where $\mu > 0$ is an algorithm parameter that encourages
sparsity in the solution $\xbfhat$.
The nonzero components of $\xbfhat$ can then be used as an estimate of $\Itrue$.

Wainwright~\cite{Wainwright:09-lasso} has given necessary and sufficient
conditions for asymptotic reliable detection with lasso.
Partly because of freedom in the choice of a sequence of parameters $\mu(n)$,
the finite SNR results are difficult to interpret.
Under certain conditions with SNR growing unboundedly with $n$,
matching necessary and sufficient conditions can be found.
Specifically,
if $m$, $n$ and $k \arr \infty$, with $\SNR \cdot \MAR \arr \infty$,
the scaling
\beq \label{eq:minMeasLasso}
    m \ > \ 2 k \log(n-k) + k + 1
\eeq
is both necessary and sufficient for asymptotic reliable detection.

Another common approach to support detection is the
OMP algorithm~\cite{ChenBL:89,PatiRK:93,DavisMZ:94}.
This was analyzed by Tropp and Gilbert~\cite{TroppG:07}
in a setting with no noise.
This was generalized to the present setting with noise by
Fletcher and Rangan~\cite{FletcherR:09}.
The result is very similar to condition (\ref{eq:minMeasLasso}):
If $m$, $n$ and $k \arr \infty$, with $\SNR \cdot \MAR \arr \infty$,
a \emph{sufficient} condition for asymptotic reliable recovery is
\beq \label{eq:minMeasOMP}
    m \ > \ 2 k\log(n-k).
\eeq
The main result of~\cite{FletcherR:09} also allows uncertainty in $k$.

The conditions (\ref{eq:minMeasLasso}) and (\ref{eq:minMeasOMP})
are both shown in Table~\ref{table:summary}.
As usual, the table entries are simplified by including only
the leading terms.

The lasso and OMP scaling laws,
(\ref{eq:minMeasLasso}) and (\ref{eq:minMeasOMP}),
can be compared with the high SNR limit for the thresholding
scaling law in (\ref{eq:minMeasSUDlim}).
This comparison shows the following:

\begin{itemize}
\item \emph{Removal of the constant offset:}
The $L(k,n)$ factor in the thresholding expression
is replaced by a $\log(n-k)$ factor in the lasso and OMP scaling laws.
Similar to the discussion above, this implies that lasso and OMP
could require up to 4 times fewer measurements than thresholding.

\item \emph{Dynamic range:}
In addition, both the lasso and OMP methods do not have a dependence
on MAR\@.
This gain can be large when there is high dynamic range, i.e., MAR is near zero.

\item \emph{Limits at high SNR:}
We also see from (\ref{eq:minMeasLasso}) and
(\ref{eq:minMeasOMP}) that both lasso and OMP are unable to achieve
the scaling $m = O(k)$ that may be achievable with ML at high SNR\@.
Instead, both lasso and OMP have the scaling $m = O(k \log(n-k))$,
similar to the minimum scaling possible with thresholding.
\end{itemize}

\subsection{Other Sparsity Detection Algorithms}
Recent interest in compressed sensing has led to a plethora
of algorithms beyond OMP and lasso.
Empirical evidence suggests that the most promising algorithms
for support detection are the sparse Bayesian learning methods
developed in the machine learning community~\cite{Tipping:01}
and introduced into signal processing applications in \cite{WipfR:04},
with related work in \cite{SchniterPZ:0x}.
Unfortunately, a comprehensive summary of these
algorithms is far beyond the scope of
this paper.
Our interest is not in finding the optimal algorithm,
but rather to explain qualitative differences between algorithms
and to demonstrate the value of knowing conditional ranks
\emph{a priori}.

\section{Sequential Orthogonal Matching Pursuit}
\label{sec:SOMP}

The results summarized in the previous section suggest a large performance gap
between ML detection and practical algorithms such as thresholding,
lasso and OMP, especially when the SNR is high.
Specifically, as the SNR increases, the performance of these practical
methods saturates at a scaling in the number of measurements that
can be significantly higher than that for ML\@.

In this section, we introduce an OMP-like algorithm, which we call
\emph{sequential orthogonal matching pursuit}, that under favorable
conditions can break this barrier.
Specifically, in some cases,
the performance of SequOMP does not saturate at high SNR.

\subsection{Algorithm: SequOMP} \label{sect:somp}
Given a received vector $\ybf$,
threshold level $\mu > 0$, and detection order $\pi$
(a permutation on $\{1,\,2,\,\ldots,\,n\}$),
the algorithm produces an estimate $\IhatSOMP$ of the support $\Itrue$
with the following steps:
\begin{enumerate}
\item Initialize the counter $j=1$ and
set the initial support estimate to empty:  $\Ihat(0) = \{\emptyset\}$.
\item Compute $\Pbf(j)\abf_{\pi(j)}$
where $\Pbf(j)$ is the projection operator
onto the orthogonal complement
of the span of $\{ \abf_{\pi(\ell)},\, \pi(\ell) \in \Ihat(j-1)\}$.
\item Compute the correlation
$$
    \rho(j) \ = \ \frac{|\abf_{\pi(j)}'\Pbf(j)\ybf|^2 }
                   {\|\Pbf(j)\abf_{\pi(j)}\|^2\|\Pbf(j)\ybf\|^2}.
$$
\item If $\rho(j) > \mu$, add the index $\pi(j)$ to $\Ihat(j-1)$.  That is,
$\Ihat(j) = \Ihat(j-1) \cup \{j\}$.
Otherwise, set $\Ihat(j) = \Ihat(j-1)$.
\item Increment $j = j +1$.  If $j \leq n$ return to step 2.
\item The final estimate of the support is $\IhatSOMP = \Ihat(n)$.
\end{enumerate}

\smallskip

The SequOMP algorithm can be thought of as an iterative version of thresholding
with the difference that, after a nonzero component
is detected, subsequent correlations are performed only in the orthogonal complement
to the corresponding column of $\Abf$.  The method
is identical to the standard OMP algorithm of
\cite{ChenBL:89,PatiRK:93,DavisMZ:94},
except that SequOMP passes through the data only once,
in a fixed order.
For this reason, SequOMP is computationally simpler than standard OMP\@.

As simulations will illustrate later, SequOMP generally has much worse
performance than standard OMP\@.  It is not intended as a competitive practical
alternative.  Our interest in the algorithm lies in the fact
that we can prove positive results for SequOMP\@.
Specifically, we will be able to show that this simple algorithm,
when used in conjunction with known conditional ranks,
can achieve a fundamentally better scaling at high SNRs
than what has been proven is achievable with methods such as lasso and OMP\@.

\subsection{Sequential OMP Performance}
The analyses in Section~\ref{sec:csAnalysis} hold for deterministic
vectors $\xbf$.
Recall the partially-random signal model (\ref{eq:xmodel}) where $b_j$
is a Bernoulli($\lambda$) random variable while the value of $x_j$ conditional on
$x_j$ being nonzero remains deterministic; i.e., $s_j$ is deterministic.

Let $p_j$ denote the conditional energy of $x_j$, conditioned on $b_j = 1$
(i.e., $j \in \Itrue$).  Then
\beq \label{eq:pjdef}
    p_j \ = \ s_j^2,
\qquad j = 1,\,2,\,\ldots,\,n.
\eeq
We will call $\{p_j\}_{j=1}^n$ the \emph{power profile}.
Since $\Pr(b_j = 1) = \lambda$ for every $j$, the average value of $\SNR(\xbf)$
in (\ref{eq:snrDef}) is given by
\beq \label{eq:snrp}
    \SNR = \lambda \sum_{j=1}^n p_j.
\eeq
Also, in analogy with $\MAR(\xbf)$ and $\SNRmin(\xbf)$
in (\ref{eq:MAR-def}) and (\ref{eq:snrMinDef}), define
\beqan
    \SNRmin &=& \min_j p_j, \label{eq:snrMinp} \\
    \MAR &=& \frac{\lambda n}{\SNR}\min_j p_j
    \ = \  \frac{\lambda n \, \SNRmin}{\SNR} \label{eq:MARp}.
\eeqan
Note that the power profile $p_j$ and the quantities $\SNR$, $\SNRmin$
and $\MAR$ as defined above are deterministic.

To simplify notation, we henceforth assume $\pi$ is the identity permutation,
i.e., the detection order in SequOMP is simply $(1,\,2,\,\ldots,\,n)$.
A key parameter in analyzing the performance of SequOMP
is what we will call
the \emph{minimum signal-to-interference and noise ratio (MSINR)}
\beq \label{eq:sinrDef}
    \gamma \ = \ \min_{\ell=1,\ldots,n}  p_\ell / \sigmahat^2(\ell),
\eeq
where $\sigmahat^2(\ell)$ is given by
\beq \label{eq:sigHatDef}
    \sigmahat^2(\ell) \ = \ 1 + \lambda \sum_{j=\ell+1}^n p_j,
\qquad \ell = 1,\,2,\,\ldots,\,n.
\eeq
The parameters $\gamma$ and $\sigmahat^2(\ell)$ have simple interpretations:
Suppose SequOMP
has correctly detected $b_j$ for all $j < \ell$.
Then, in detecting $b_\ell$, the algorithm sees the noise $\dbf$
with power $\Exp[\|\dbf\|^2] = 1$ plus, for each component $j > \ell$,
an interference power $p_j$ with probability $\lambda$.
Hence, $\sigmahat^2(\ell)$ is the total average interference power seen when
detecting $b_\ell$, assuming perfect cancellation up to that point.
Since the conditional power of $x_\ell$ is $p_\ell$,
the ratio $p_\ell / \sigmahat^2(\ell)$
in (\ref{eq:sinrDef}) represents the average SINR seen while detecting
component $\ell$.
The value $\gamma$ is the minimum SINR over all $n$ components.

\medskip

\begin{theorem} \label{thm:minMeasSOMP}
Let $\lambda = \lambda(n)$, $m = m(n)$, and
the power profile $\left\{p_j\right\}_{j=1}^n = \left\{p_j(n)\right\}_{j=1}^n$
be deterministic quantities that all vary with $n$
satisfying the limits
$$
m-\lambda n \rightarrow \infty,
  \quad
\lambda n \rightarrow \infty,
  \quad
(1- \lambda) n \arr \infty,
  \quad
\mbox{and}
  \quad
\gamma \arr 0.
$$
Also, assume the sequence of power profiles satisfies the limit
\beq \label{eq:pjsqbnd}
    \limn \max_{i=1,\ldots,n-1} \log(n)\sigmahat^{-4}(i)\sum_{j>i}^n p_j^2 \ = \ 0.
\eeq
Finally, assume that for all $n$,
\beq \label{eq:minMeasGam}
    m \ \geq \ \frac{2(1 + \delta)L(n,\lambda)}{\gamma} + \lambda n,
\eeq
for some $\delta > 0$ and $L(n,\lambda)$ defined in (\ref{eq:Ldef}).
Then, there exists a sequence of thresholds, $\mu = \mu(n)$,
such that SequOMP will achieve asymptotic reliable detection.
The sequence of threshold levels can be selected independent of the
sequence of power profiles.
\end{theorem}
\begin{IEEEproof} See Appendix~\ref{sec:proofOutline}.
\end{IEEEproof}

The theorem provides a simple sufficient condition on the number of
measurements as a function of the MSINR $\gamma$, probability $\lambda$,
and dimension $n$.  The condition (\ref{eq:pjsqbnd})
is somewhat technical; we will verify its validity in examples.
The remainder of this section discusses some of the implications of this theorem.

\subsection{Most Favorable Detection Order with Known Conditional Ranks}
\label{sec:nearFarSOMP}

Suppose that the \emph{ordering} of the conditional power
levels $\{p_j\}_{j=1}^n$ is known at the detector,
but possibly not the values themselves.
Reordering the power profile is equivalent to changing the detection order,
so we seek the most favorable ordering of the power profile.
Since $\sigmahat^2(\ell)$ defined in (\ref{eq:sigHatDef}) involves the
sum of the tail of the power profile,
the MSINR defined in (\ref{eq:sinrDef}) is maximized when the power
profile is non-increasing:
\beq \label{eq:porder}
    p_1 \ \geq \ p_2 \ \geq \ \cdots \ \geq \ p_n \ = \ \SNRmin.
\eeq
In other words, the best detection order for SequOMP is from
strongest component to weakest component.

Using (\ref{eq:porder}),
it can be verified that the MSINR $\gamma$ is bounded below by
\beq \label{eq:gammaLow}
    \gamma \ \geq \ \frac{\SNRmin}{1+\lambda n\SNRmin}
    \ = \ \frac{\SNR \cdot \MAR}{\lambda n(1+\SNR\cdot \MAR)}.
\eeq
Furthermore, the sufficiency of the scaling (\ref{eq:minMeasGam})
shows that
\beq \label{eq:minMeasOrderFinite}
    m \ \geq \ \frac{2(1 + \delta)\lambda n (1+\SNR\cdot\MAR)}{\SNR \cdot\MAR}
        L(n,\lambda) + \lambda n
\eeq
is sufficient for asymptotic reliable detection.
This expression is shown in Table \ref{table:summary} with the
additional simplification that $L(n,\lambda) \leq 4\log(n(1-\lambda))$
for $\lambda \in (0,1/2)$.
To keep the notation consistent with the expressions for the
other entries in the table, we have used $k$ for $\lambda n$, which is
the average number of non-zero entries of $\xbf$.

When $\SNR \arr \infty$, (\ref{eq:minMeasOrderFinite}) simplifies to
\beq \label{eq:minMeasOrder}
    m \ \geq \ 2(1 + \delta)\lambda n L(n,\lambda) + \lambda n.
\eeq
This is identical to the lasso and OMP performance except for the factor
$L(\lambda,n) / \log((1-\lambda)n)$, which lies in $(0,4)$ for
$\lambda \in (0,1/2)$.
In particular, the minimum number of measurements does not depend on $\MAR$;
therefore, similar to lasso and OMP, SequOMP can theoretically detect
components that are much below the average power at high SNRs.
More generally, we can say that knowledge of the conditional
ranks of the powers enable a very simple algorithm to achieve resistance
to large dynamic ranges.

\subsection{Optimal Power Shaping}
\label{sec:powShaping}

The MSINR lower bound in (\ref{eq:gammaLow}) is achieved as $n \arr\infty$
and the power profile is constant (all $p_j$'s are equal).  Thus, opposite to
thresholding, a constant power profile is in some sense the \emph{worst}
power profile for a given $\SNRmin$ for the SequOMP algorithm.

This raises the question: What is the most favorable power profile?
Any power profile maximizing the MSINR $\gamma$
subject to a constraint on total SNR (\ref{eq:snrp})
will achieve the minimum in (\ref{eq:sinrDef}) for every $\ell$
and thus satisfy
\beq \label{eq:pellSnrCon}
    p_\ell \ = \ \gamma\Bigg( 1 + \lambda \sum_{j=\ell+1}^n p_j \Bigg),
\qquad \ell = 1,\,2,\,\ldots,\,n.
\eeq
The solution to (\ref{eq:pellSnrCon}) and (\ref{eq:snrp})
is given by
\begin{subequations}
\label{eqs:powerProfileNoLeakage}
\beq \label{eq:pexp}
    p_\ell \ = \ \gammaOpt(1 + \gammaOpt \lambda)^{n-\ell},
\qquad \ell = 1,\,2,\,\ldots,\,n,
\eeq
where
\beq \label{eq:gamExp}
    \gammaOpt \ = \ \frac{1}{\lambda}\left[(1 + \SNR)^{1/n} - 1\right]
    \ \approx \ \frac{1}{\lambda n}\log(1+\SNR)
\eeq
\end{subequations}
and the approximation holds for large $n$.\footnote{The solution
(\ref{eqs:powerProfileNoLeakage}) is the $\theta = 0$ case of a
more general result in Section~\ref{sec:robustPowerShaping};
see (\ref{eqs:powerProfileLeakage}).}
Again, some algebra shows that when $\lambda$ is bounded away from zero,
the power profile
in (\ref{eqs:powerProfileNoLeakage}) will
satisfy the technical condition (\ref{eq:pjsqbnd}) when $\log(1+\SNR) = o(n/\log(n))$.

The power profile (\ref{eq:pexp}) is exponentially decreasing in the index order $\ell$.
Thus, components early in the detection sequence are allocated exponentially higher
power than components later in the sequence.  This allocation insures that
early components have sufficient power to overcome the interference
from all the components later in the detection sequence that are not yet cancelled.

Substituting (\ref{eq:gamExp}) into (\ref{eq:minMeasGam}), we see that
the scaling
\beq \label{eq:minMeasSOMP}
    m \ \geq \ \frac{2(1 + \delta)L(n,\lambda)}{\log(1+\SNR)}\lambda n + \lambda n
\eeq
is sufficient for SequOMP to achieve asymptotic reliable detection
with the best-case power profile.
This expression is shown in Table \ref{table:summary},
again with the
additional simplification that $L(n,\lambda) \leq 4\log(n(1-\lambda))$
for $\lambda \in (0,1/2)$.

\subsection{SNR Saturation}
\label{sec:snrSaturation}
As discussed earlier, a major problem with thresholding, lasso, and OMP
is that their performances ``saturate'' with high SNR\@.
That is, even as the SNR scales to infinity, the minimum number of measurements
scales as $m = \Theta(\lambda n\log((1-\lambda)n)$.
In contrast, optimal ML detection can achieve a scaling $m = O(\lambda n)$,
when the SNR is sufficiently high.

A consequence of (\ref{eq:minMeasSOMP}) is that SequOMP
with exponential power shaping can overcome this barrier.
Specifically, if we take the scaling of $\SNR = \Theta(\lambda n)$ in
(\ref{eq:minMeasSOMP}), apply the bound
$L(n,\lambda) \leq 4\log(n(1-\lambda))$ for $\lambda \in (0,1/2)$,
and assume that $\lambda$ is bounded away from zero,
we see that asymptotically, SequOMP requires only
\beq \label{eq:minMeasSMPHigh}
    m \ \geq \ 9\lambda n
\eeq
measurements.
In this way, unlike thresholding and lasso,
SequOMP is able to succeed with scaling $m = O(\lambda n)$ when
$\SNR \arr \infty$.

\subsection{Power Shaping with Sparse Bayesian Learning}
The fact that power shaping can provide benefits when
combined with certain iterative detection algorithms confirms the
observations in the work of Wipf and Rao~\cite{WipfR:06}.
That work considers signal detection with
a certain sparse Bayesian learning (SBL) algorithm.
They show the following result: Suppose $\xbf$ has $k$ nonzero
components and $p_i$, $i=1,2,\ldots,k$,
is the power of the $i$th largest component.
Then, for a given measurement matrix $\Abf$,
there exist constants $\nu_i > 1$ such that if
\beq \label{eq:sblPow}
    p_i \ \geq \ \nu_i p_{i-1},
\qquad i = 2,\,3,\,\ldots,\,k,
\eeq
the SBL algorithm will correctly detect the sparsity pattern of $\xbf$.

The condition (\ref{eq:sblPow}) shows that a certain
growth in the powers can guarantee correct detection.
The parameters $\nu_i$ however depend in some complex manner on
the matrix $\Abf$,
so the appropriate growth is difficult to compute.
They also provide strong empirical evidence that shaping
the power with certain profiles can greatly reduce the number of
measurements needed.

The results in this paper add to Wipf and Rao's observations
showing that growth in the powers can also assist SequOMP\@.
Moreover, for SequOMP, we can explicitly derive the
optimal power profile for certain large random matrices.

This is not to say that SequOMP is better than SBL\@.  In fact,
empirical results in \cite{WipfR:04} suggest that SBL will outperform
OMP, which will in turn do better than SequOMP\@.  As we have stressed before,
the point of analyzing SequOMP here is that we can derive
concrete analytic results.  These results may provide guidance for
more sophisticated algorithms.

\subsection{Robust Power Shaping}
\label{sec:robustPowerShaping}
The above analysis shows certain benefits of SequOMP used in conjunction with power shaping.
However, these gains are theoretically only possible at infinite block lengths.
Unfortunately, when the block length is finite, power shaping can actually
reduce the performance.

The problem is that when a nonzero component is not detected in SequOMP,
that component's energy is not cancelled out and remains as
interference for all subsequent components in the detection sequence.
With power shaping, components early in the detection sequence have much higher
power than components later in the sequence, so an early missed detection
can make subsequent detection difficult.  As block length increases,
the probability of missed detection can be driven to zero.  But
at any finite block length, the probability of a missed detection early
in the sequence will always be nonzero.

The work \cite{AgrawalACM:05} observed a similar problem when successive
interference cancellation is used in a CDMA uplink.
To mitigate the problem, \cite{AgrawalACM:05} proposed to adjust
the power allocations to make them more robust to detection errors early in the
detection sequence.
The same technique, which we will call \emph{robust power shaping},
can be applied to SequOMP as follows.

The condition (\ref{eq:pellSnrCon}) is motivated by maintaining a constant
MSINR through the detection process, assuming all
components with indexes $j < \ell$ have been correctly detected and subtracted.
An alternative, following \cite{AgrawalACM:05}, is to assume that some
fixed fraction $\theta \in [0,1]$ of the energy
of components early in the detection sequence is not cancelled out due to
missed detections.  We will call $\theta$ the \emph{leakage fraction}.
With nonzero leakage, the condition (\ref{eq:pellSnrCon})
is replaced by
\beq \label{eq:pellSnrConTheta}
    p_\ell \ = \ \gamma \Bigg( 1 + \theta \lambda \sum_{j=1}^{\ell-1} p_j
        +\lambda \sum_{j=\ell+1}^n p_j \Bigg),
\qquad \ell = 1,\,2,\,\ldots,\,n.
\eeq
For given $\gamma$, $\lambda$, and $\theta$,
(\ref{eq:pellSnrConTheta}) in a system of linear equations that determine
the power profile $\{p_\ell\}_{\ell=1}^n$;
one can vary $\gamma$ until the power profile provides the desired
SNR according to (\ref{eq:snrp}).

A closed-form solution to (\ref{eq:pellSnrConTheta})
provides some additional insight.
Adding and subtracting $\SNR$ inside the parentheses in
(\ref{eq:pellSnrConTheta}) while also using (\ref{eq:snrp}) yields
$$
    p_\ell \ = \ \gamma \Bigg( 1 +
      \underbrace{\SNR - \lambda \sum_{j=1}^n p_j}_{=0}
                                 + \theta \lambda \sum_{j=1}^{\ell-1} p_j
                                 + \lambda \sum_{j=\ell+1}^n p_j \Bigg),
$$
which can be rearranged to
\beq \label{eq:powerLeakageSolution1}
  (1+\gamma \lambda)p_\ell
    \ = \ \gamma \Bigg( 1 + \SNR - (1-\theta) \lambda \sum_{j=1}^{\ell -1} p_j \Bigg).
\eeq
Using standard techniques for solving linear constant-coefficient
difference equations,
\begin{subequations}
\label{eqs:powerProfileLeakage}
\beq \label{eq:powTheta}
  p_j \ = \ \frac{\SNR}{\lambda} \cdot
            \frac{(1 - \zeta) \zeta^{j-1}}{1 - \zeta^n}
\eeq
where
\beq \label{eq:optZetaGamma}
  \zeta \ = \ \frac{1 + \gamma \theta \lambda}{1 + \gamma \lambda}
\eeq
and
\beq
  \label{eq:gamTheta}
  \gamma \ = \ \frac{1}{\lambda} \cdot
     \frac{1-\left(\frac{1+\theta\,\tableSNR}{1+\tableSNR}\right)^{1/n}}
          {\left(\frac{1+\theta\,\tableSNR}{1+\tableSNR}\right)^{1/n} -\theta}.
\eeq
\end{subequations}
Notice that $\theta < 1$ implies $\zeta < 1$,
so the power profile (\ref{eq:powTheta}) is decreasing as in
the case without leakage in Section~\ref{sec:powShaping}.
Setting $\theta = 0$ recovers (\ref{eqs:powerProfileNoLeakage}).

\section{Numerical Simulation}
\label{sec:sim}

\subsection{Threshold Settings}
The performances of the thresholding and SequOMP algorithms
depend on the setting of the threshold level $\mu$.
In the theoretical analysis of Theorem~\ref{thm:minMeasSOMP},
an ideal threshold is calculated for the limit of infinite
block length, which guarantees perfect detection of the support.
In simulations with finite block lengths,
it is more reasonable to set the
threshold based on a desired \emph{false alarm probability}.
A false alarm is the event that the algorithm falsely detects that a
component is nonzero when it is not.  For the thresholding algorithm in
Section~\ref{sect:sud} or the SequOMP algorithm in Section~\ref{sect:somp},
the false alarm probability is
\beqan
    \PFA &=& \Pr\left( j \in \Ihat \mid j \not \in \Itrue \right) \\
    &=& \Pr\left( \rho(j) > \mu \mid j \not \in \Itrue \right),
\eeqan
which is the probability that the correlation $\rho(j)$ exceeds the threshold $\mu$
when $b_j = 0$.

In the simulations below, we adjust the threshold $\mu$
by trial and error to achieve a fixed  false
alarm probability (typically $\PFA = 10^{-3}$), and then measure
the \emph{missed detection probability} given by
\[
    \PMD = \Pr\left( j \not \in \Ihat \mid j \in \Itrue \right).
\]
The missed detection probability is averaged over all $j \in \Itrue$.

\subsection{Evaluation of Bounds}
\begin{figure}
 \begin{center}
  \epsfig{figure=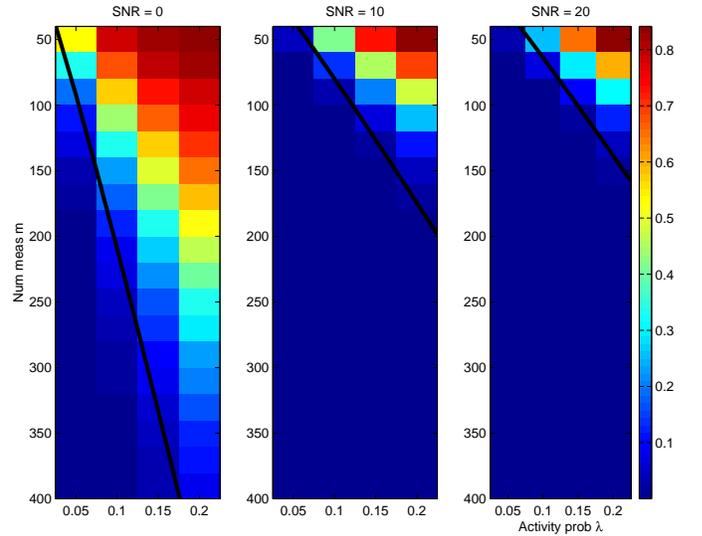,width=3.5in}
 \end{center}
 \caption{SequOMP with power shaping:
 Each colored bar represents
 the SequOMP algorithm's missed detection probability
 as a function of the number of measurements $m$,
 with different bars showing different activity probabilities $\lambda$ and
 SNR levels.  The missed detection probabilities were estimated with 1000 Monte Carlo
 trials.  The number of users is set to $n=100$, the false alarm
 probability is  $\PFA = 10^{-3}$.
 The power shaping is performed with a leakage fraction of $\theta=0.1$.
 The dark black line shows the theoretical number of measurements
 $m$ required in Theorem~\ref{thm:minMeasSOMP} with $\gamma = \gamma(\theta)$
 in (\ref{eq:gamTheta}). }
 \label{fig:sompBnd}
\end{figure}

We first compare the actual performance of the SequOMP algorithm
with the bound in Theorem~\ref{thm:minMeasSOMP}.
Fig.~\ref{fig:sompBnd} plots the simulated missed detection probability for
using SequOMP at various SNR levels, probabilities of nonzero components
$\lambda$, and numbers of measurements $m$.  In all these simulations,
the number of components was fixed to $n=100$.
The false alarm probability was set to $\PFA = 10^{-3}$.
The robust power profile of Section
\ref{sec:robustPowerShaping} is used with a leakage fraction $\theta = 0.1$.

The dark line in Fig.~\ref{fig:sompBnd} represents the number of measurements
$m$ for which Theorem~\ref{thm:minMeasSOMP} would theoretically guarantee reliable
detection of the support at infinite block lengths.
To apply the theorem, we used the MSINR $\gamma = \gamma(\theta)$ in (\ref{eq:gamTheta}).
At the block lengths considered in this simulation,
the missed detection probability at the theoretical sufficient condition is small,
typically between 2 and 10\%.
Thus, even at moderate block lengths, the theoretical bound
in Theorem~\ref{thm:minMeasSOMP} can provide a good estimate for the number of
measurements for reliable detection.

\subsection{SequOMP vs.\ Thresholding}

\begin{figure}
 \begin{center}
  \epsfig{figure=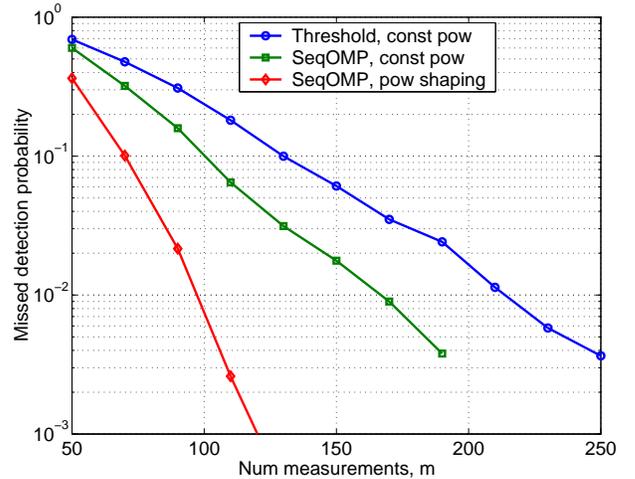,width=3.2in}
 \end{center}
 \caption{Missed detection probabilities for various
  detection methods and power profiles.
  The number of users is $n=100$, $\captionSNR = 20\,\dB$,
  the activity probability is $\lambda = 0.1$, and
  the false alarm rate is $\PFA = 10^{-3}$.
  For the SequOMP algorithm with power shaping,
   the leakage fraction was set to $\theta = 0.1$.}
 \label{fig:sompSim}
\end{figure}

Fig.~\ref{fig:sompSim} compares the performances of thresholding
and SequOMP with power shaping.
In the simulations, $n =100$, $\lambda = 0.1$,
and the total SNR is 20 dB\@.
The number of measurements $m$ was varied, and for each $m$,
the missed detection probability was estimated with 1000 Monte Carlo
trials.

As expected, thresholding requires the most number of measurements.
For a missed detection rate of 1\%, Fig.~\ref{fig:sompSim} shows that
thresholding requires approximately $m \approx 210$ measurements.
In this simulation of thresholding, the power profile is constant.
Employing SequOMP but keeping the power profile constant
decreases the number of measurements somewhat to $m \approx 170$ for
a 1\% missed detection rate.
However, using SequOMP with power shaping decreases the number of measurements
by more than a factor of two to $m \approx 95$.
Thus, at least at high SNRs, SequOMP may provide significant gains over simple
thresholding.

\subsection{OMP with Power Shaping}

As discussed earlier, although SequOMP can provide gains over thresholding,
its performance is typically worse than OMP,
even if SequOMP is used with power shaping.
(Our interest in SequOMP is that it is simple to analyze.)

\begin{figure}
 \begin{center}
  \epsfig{figure=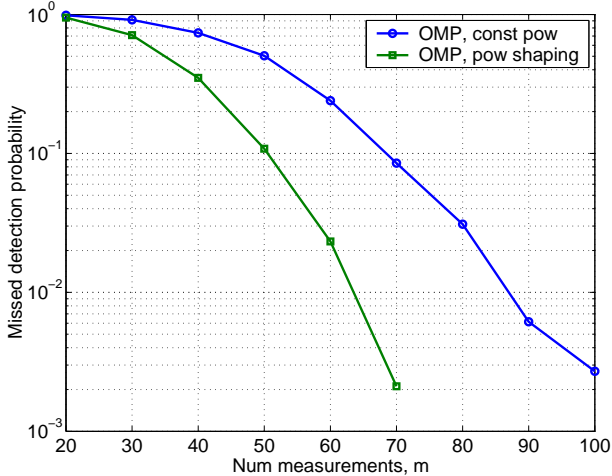,width=3.2in}
 \end{center}
 \caption{Power shaping with OMP\@.  Plotted is the missed detection
 probabilities with OMP using a constant power profile, and
 power shaping wiht a leakage fraction set to $\theta = 0.1$.
 Other simulation assumptions are identical to Fig.~\ref{fig:sompSim}.}
 \label{fig:ompSim}
\end{figure}

While we do not have any analytical result,
the simulation in Fig.~\ref{fig:ompSim}
shows that power shaping provides gains with OMP as well.
Specifically, when the power profile is constant,
$m \approx 85$ measurements are needed for a missed detection probability of 1\%.
This number is slightly lower than that required by SequOMP,
even when SequOMP uses power shaping.
When OMP is used with power shaping,
the number of measurements decreases to $m \approx 65$.

\section{Conclusions}
\label{sec:concl}
Methods such as OMP and lasso, which are widely used in sparse signal
support detection problems, exhibit advantages over thresholding but
still fall far short of the performance of optimal (ML) detection at
high SNRs.
Analysis of the SequOMP algorithm has shown that knowledge of conditional
rank of signal components enables performance similar to OMP and lasso
at a lower complexity.  Furthermore, in the most favorable situations,
conditional rank knowledge changes the fundamental scaling of performance
with SNR so that performance no longer saturates with SNR\@.

\appendix

\section*{Proof of Theorem~\ref{thm:minMeasSOMP} }
\label{sec:proofs}

\subsection{Proof Outline}
\label{sec:proofOutline}
At a high level, the proof of Theorem~\ref{thm:minMeasSOMP} is similar
to the proof of \cite[Thm.~2]{FletcherRG:09-IT},
the thresholding condition (\ref{eq:minMeasSUD}).
One of the difficulties in the proof is to handle the dependence
between random events at different iterations of the SequOMP algorithm.
To avoid this difficulty, we first show an equivalence
between the success of SequOMP and an alternative sequence of events that
is easier to analyze.
After this simplification,
small modifications handle the cancellations of detected vectors.

Fix $n$ and define
\[
    \Itrue(j) = \left\{ ~\ell~: ~\ell \in \Itrue, \ell \leq j \right\},
\]
which is the set of elements of the true support with indices $\ell \leq j$.
Observe that $\Itrue(0) = \{\emptyset\}$ and $\Itrue(n) = \Itrue$.

Let $\Ptrue(j)$ be the projection operator onto the orthogonal complement
of $\{ \abf_\ell, ~\ell \in \Itrue(j-1)\}$, and define
\beq \label{eq:rhotrue}
    \rhotrue(j) = \frac{|\abf_j'\Ptrue(j)\ybf|^2}{\|\Ptrue(j)\abf_j\|^2\|\Ptrue(j)\ybf\|^2}.
\eeq
A simple induction argument shows that SequOMP
correctly detects the support if and only if, at each
iteration $j$, the variables $\Ihat(j)$, $\Pbf(j)$ and $\rho(j)$ defined
in the algorithm are equal to $\Itrue(j)$, $\Ptrue(j)$ and $\rhotrue(j)$,
respectively.
Therefore, if we define
\beq \label{eq:IhatSOMPtrue}
    \Ihat = \left\{ ~j~: ~\rhotrue(j) > \mu ~\right\},
\eeq
then
SequOMP
correctly detects the support if
and only if $\Ihat = \Itrue$.
In particular,
\[
    \Perr(n) = \Pr\left( \Ihat \neq \Itrue \right).
\]

To prove that $\Perr(n) \arr 0$ it suffices to show that
there exists a sequence of threshold levels $\mu(n)$
such the following two limits
\beqa
    \liminf_{n \arr \infty} \min_{j \in \Itrue(n)} \frac{\rhotrue(j)}{\mu} > 1,
        \label{eq:rhoMD} \\
    \limsup_{n \arr \infty} \max_{j \not \in \Itrue(n)} \frac{\rhotrue(j)}{\mu} < 1,
        \label{eq:rhoFA}
\eeqa
hold in probability.
The first limit (\ref{eq:rhoMD}) ensures that all the components in the
true support will not be missed and will be called the \emph{zero missed detection
condition}.
The second limit (\ref{eq:rhoFA}) ensures that all the components not
in the true support will not be falsely detected and will be called the
\emph{zero false alarm condition}.

Set the sequence of threshold levels as follows.
Since $\delta > 0$, we can find an $\epsilon > 0$ such that
\beq \label{eq:epsDef}
    (1+\delta) \geq (1+\epsilon)^2.
\eeq
For each $n$, let the threshold level be
\beq \label{eq:muDef}
    \mu = (1+\epsilon)\frac{\log(n(1-\lambda))}{m - \lambda n}.
\eeq
The asymptotic lack of missed detections and false alarms
with these thresholds are proven in Appendices~\ref{sec:pmd}
and~\ref{sec:pfa}, respectively.
In preparation for these sections,
Appendix~\ref{sec:chiSq} reviews some facts concerning tail
bounds on Chi-squared and Beta random variables and Appendix
\ref{sec:prelimComp} performs some preliminary computations.

\subsection{Chi-Squared and Beta Random Variables}
\label{sec:chiSq}
The proof requires a number of simple facts concerning
chi-squared and beta random variables.
These variables are reviewed in~\cite{EvansHP:00}.
We will omit all the proofs in this subsections as they can be
proved along the lines of the calculations in \cite{FletcherRG:09-IT}.

A random variable $u$ has a \emph{chi-squared} distribution with $r$
degrees of freedom if it can be written as
    $u = \sum_{i=1}^r z_i^2$,
where $z_i$ are i.i.d.\ ${\mathcal{N}}(0,1)$.

\begin{lemma} \label{lem:chiSqDef}
Suppose $\xbf \in \R^r$ has a Gaussian distribution
${\mathcal{N}}(0,\sigma^2I_r)$.  Then:
\begin{itemize}
\item[(a)] $\|\xbf\|^2/\sigma^2$ is chi-squared with $r$ degrees of freedom; and
\item[(b)] if $\ybf$ is any other $r$-dimensional random vector that is
nonzero with probability one and independent of $\xbf$, then the variable
\[
    u = \frac{|\xbf'\ybf|^2}{\sigma^2\|\ybf\|^2}
\]
is a chi-squared random variable with one degree of freedom.
\end{itemize}
\end{lemma}

The following two lemmas provide standard tail bounds.

\begin{lemma} \label{lem:chiSqEq}
Suppose that for each $n$, $\{\xbf_j^{(n)}\}_{j=1}^n$
is a set of Gaussian random vectors with each $\xbf_j^{(n)}$
spherically symmetric in an $m_j(n)$-dimensional space.
The variables may be dependent.  Suppose also that $\Exp\|\xbf_j^{(n)}\|^2 = 1$
and
\[
    \limn \log(n) / \mmin(n) = 0
\]
where
\[
    \mmin(n) = \min_{j=1,\ldots,n} m_j(n).
\]
Then the limits
\[
    \limn \max_{j=1,\ldots,n} \|\xbf_j^{(n)}\|^2
    = \limn \min_{j=1,\ldots,n}  \|\xbf_j^{(n)}\|^2 = 1
\]
hold in probability.
\end{lemma}

\begin{lemma} \label{lem:chiSqMax}  Suppose that for each $n$,
$\{u_j^{(n)}\}_{j=1}^n$ is a set of chi-squared random variables,
each with one degree of freedom.
The variables may be dependent.
Then
\beq \label{eq:chiSqMaxa}
    \limsupn \max_{j=1,\ldots,n} \frac{u_j^{(n)}}{2\log(n)} \leq 1 ,
\eeq
where the limit is in probability.
\end{lemma}

The final two lemmas concern certain beta distributed random variables.
A real-valued scalar random variable $w$ follows
a $\BetaDist(r,s)$ distribution if it can be written as
    $w = u_r / (u_r + v_s)$,
where the variables $u_r$ and $v_s$ are independent chi-squared random variables
with $r$ and $s$ degrees of freedom, respectively.  The importance of the
beta distribution is given by the following lemma.

\begin{lemma} \label{lem:betaProj}
Suppose $\xbf$ and $\ybf$ are independent random $r$-dimensional
random vectors with $\xbf$ being spherically-symmetrically distributed
in $\R^r$ and $\ybf$ having any distribution
that is nonzero with probability one.  Then the random variable
\[
    w = \frac{|\xbf'\ybf|^2}{\|\xbf\|^2\|\ybf\|^2}
\]
is independent of $\xbf$ and follows a $\BetaDist(1,r-1)$ distribution.
\end{lemma}

The following lemma provides a simple expression for the
maxima of certain beta distributed variables.

\begin{lemma} \label{lem:betaLim}
For each $n$, suppose $\{w_j^{(n)}\}_{j=1}^n$
is a set of random variables with $w_j^{(n)}$ having a
$\BetaDist(1,m_j(n)-1)$ distribution.
Suppose that
\beq \label{eq:limlogb}
    \limn \log(n) / \mmin(n) = 0, \ \ \
    \limn \mmin(n) = \infty
\eeq
where
\[
    \mmin(n) = \min_{j=1,\ldots,n} m_j(n).
\]
Then,
\[
    \limsupn \max_{j=1,\ldots,n} \frac{m_j(n)}{2\log(n)}w_j^{(n)} \leq 1
\]
in probability.
\end{lemma}

\subsection{Preliminary Computations and Technical Lemmas}
\label{sec:prelimComp}
We first need to prove a number of simple but technical bounds.
We begin by considering the dimension $m_i$ defined as
\beq \label{eq:miDef}
    m_i = \dim(\range(\Ptrue(i))).
\eeq
Our first lemma computes the limit of this dimension.

\begin{lemma} \label{lem:miMin}
The following limit
\beq \label{eq:miMin}
    \lim_{n \arr \infty} \min_{i=1,\ldots,n} \frac{m_i}{m - \lambda n} = 1
\eeq
holds in probability and almost surely.  The deterministic limits
\beq \label{eq:lognm}
    \lim_{n \arr \infty} \frac{\log(\lambda n)}{m - \lambda n}
    = \lim_{n \arr \infty} \frac{\log((1-\lambda) n)}{m - \lambda n} = 0
\eeq
also hold.
\end{lemma}
\begin{IEEEproof}
Recall that $\Ptrue(i)$ is the projection onto the orthogonal complement
of the vectors $\abf_j$ with $j \in \Itrue(i-1)$.
With probability one, these vectors will be linearly independent, so $\Ptrue(i)$
will have dimension $m - |\Itrue(i-1)|$.  Since $\Itrue(i)$ is increasing with $i$,
\beqa
    \min_{i=1,\ldots,n} m_i &=& m - \max_{i=1,\ldots,n} |\Itrue(i-1)| \nonumber \\
    &=& m - |\Itrue(n-1)|. \label{eq:miItrue}
\eeqa
Since each user is active with probability $\lambda$ and the activities of the
users are independent, the law of large numbers shows that
\[
    \lim_{n \arr \infty} \frac{|\Itrue(n-1)|}{\lambda(n-1)} = 1
\]
in probability and almost surely.  Combining this with (\ref{eq:miItrue})
shows (\ref{eq:miMin}).

We next show (\ref{eq:lognm}).
Since the hypothesis of the theorem requires that
$\lambda n$, $(1-\lambda)n$ and $m -\lambda n$ all approach infinity,
the fractions in (\ref{eq:lognm}) are eventually positive.
Also, from (\ref{eq:Ldef}), $L(\lambda,n) < \max\{\log(\lambda n), \log((1-\lambda)n)\}$.
Therefore, from (\ref{eq:minMeasGam}),
\beqan
    \lefteqn{ \frac{1}{m-\lambda n}\max\{\log(\lambda n), \log((1-\lambda)n)\} }\\
    &\leq& \frac{\gamma}{2L(\lambda,n)}\max\{\log(\lambda n), \log((1-\lambda)n)\}
    \leq \frac{\gamma}{2} \arr 0,
\eeqan
where the last step is from the hypothesis of the theorem.
\end{IEEEproof}

\medskip
Next, for each $i=1,\ldots,n$, define the \emph{residual vector},
\beq \label{eq:residDef}
    \ebf_i = \Ptrue(i)(\ybf - \abf_ix_i).
\eeq
Observe that
\beqa
    \ebf_i &=& \Ptrue(i)(\ybf - \abf_ix_i) \nonumber \\
        &\stackrel{(a)}{=}& \Ptrue(i)\left(\dbf + \sum_{j \neq i} \abf_jx_j \right)
        \nonumber \\
        &\stackrel{(b)}{=}& \Ptrue(i)\left(\dbf + \sum_{j > i} \abf_jx_j \right)
        \label{eq:residSum}
\eeqa
where (a) follows from (\ref{eq:yax}) and
(b) follows from the fact that $\Ptrue(i)$ is the projection onto
the orthogonal complement of the span of all vectors $\abf_j$ with $j < i$
and $x_j \neq 0$.

The next lemma shows that the power of the residual
vector is described by the random variable
\beq\label{eq:sigDef}
    \sigma^2(i) = 1 + \sum_{j = i+1}^n |x_j|^2.
\eeq

\medskip
\begin{lemma} \label{lem:residDist}
For all $i=1,\ldots,n$, the residual vector $\ebf_i$,
conditioned on the modulation vector $\xbf$ and projection $\Ptrue(i)$,
is a spherically symmetric Gaussian in the range space of $\Ptrue(i)$
with total variance
\beq \label{eq:residPow}
    \Exp\left(\|\ebf_i\|^2 \mid \xbf \right) = \frac{m_i}{m}\sigma^2(i),
\eeq
where $m_i$ and $\sigma^2(i)$ are defined in (\ref{eq:miDef}) and (\ref{eq:sigDef}),
respectively.
\end{lemma}
\begin{IEEEproof}
Let
\[
    \vbf_i = \dbf + \sum_{j > i} \abf_jx_j,
\]
so that $\ebf_i = \Ptrue(i)\vbf_i$.
Since the vectors $\abf_j$ and $\dbf$ have Gaussian
${\mathcal{N}}(0,1/mI_m)$ distributions, for a given vector $\xbf$,
$\vbf_i$ must be a zero-mean white Gaussian vector with total variance
$\Exp\|\vbf_i\|^2 = \sigma^2(i)$.
Also, since the operator $\Ptrue(i)$ is a function of the components
$x_\ell$ and vectors $\abf_\ell$ for $\ell < i$,
$\Ptrue(i)$ is independent of the vectors $\dbf$ and $\abf_j$, $j > i$,
and therefore independent of $\vbf_i$.
Since $\Ptrue(i)$ is a projection from an $m$-dimensional space
to an $m_i$-dimensional space, $\ebf_i$, conditioned on the modulation
vector $\xbf$, must be spherically symmetric Gaussian in the range space
of $\Ptrue(i)$ with total variance satisfying (\ref{eq:residPow}).
\end{IEEEproof}

Our next lemma requires the following version of
the well-known Hoeffding's inequality.

\begin{lemma}[Hoeffding's Inequality] \label{lem:hoeffding}
Suppose $z$ is the sum
\[
    z = z_0 + \sum_{i=1}^r z_i
\]
where $z_0$ is a constant and the variables $z_i$ are independent random variables
that are almost surely bounded in some interval $z_i \in [a_i,b_i]$.
Then, for all $\epsilon > 0$,
\[
    \Pr\left( z - \Exp(z) \geq \epsilon \right) \leq
        \exp\left(\frac{-2\epsilon^2}{C}\right),
\]
where
\[
    C = \sum_{i=1}^r (b_i-a_i)^2.
\]
\end{lemma}
\begin{IEEEproof}  See \cite{Hoeffding:63}.
\end{IEEEproof}

\begin{lemma} \label{lem:sigLim}
Under the assumptions of Theorem~\ref{thm:minMeasSOMP}, the limit
\[
    \limsup_{n \arr \infty} \max_{i=1,\ldots,n} \frac{\sigma^2(i)}{\sigmahat^2(i)} \leq 1
\]
holds in probability.
\end{lemma}
\begin{IEEEproof}
Let $z(i) = \sigma^2(i)/\sigmahat^2(i)$.
From the definition of $\sigma^2(i)$ in (\ref{eq:sigDef}), we can write
\[
    z(i) = \frac{1}{\sigmahat^2(i)} + \sum_{j=i+1}^n z(i,j),
\]
where $z(i,j) = |x_j|^2/\sigmahat^2(i)$ for $j > i$.

Now recall that in the problem formulation,
each user is active with probability $\lambda$,
with power $|x_j|^2 = p_j$ conditioned on when the user being active.
Also, the activities of different users are independent,
and the conditional powers $p_j$ are treated as deterministic
quantities.
Therefore, the variables $z(i,j)$ are independent with
\[
    z(i,j) = \left\{ \begin{array}{ll}
        p_j/\sigmahat^2(i), & \mbox{with probability $\lambda$;} \\
        0,                  & \mbox{with probability $1-\lambda$},
        \end{array} \right.
\]
for $j > i$.
Combining this with the definition of $\sigmahat^2(i)$ in (\ref{eq:sigHatDef}),
we see that
\[
    \Exp(z(i)) = \frac{1}{\sigmahat^2(i)}\left(1 + \lambda \sum_{j=i+1}^n p_j\right) = 1.
\]
Also, for each $j > i$, we have the bound
\[
    z(i,j) \in [0, p_j/\sigmahat^2(i)].
\]
So for use in Hoeffding's Inequality (Lemma~\ref{lem:hoeffding}),
define
\[
    C = C(i,n) = \sigmahat^{-4}(i)\sum_{j = i+1}^n p_j^2,
\]
where dependence of the power profile and $\sigmahat(i)$ on $n$ is implicit.
Now define
\[
    c_n = \max_{i=1,\ldots,n} \log(n) C(i,n),
\]
so that $C(i,n) \leq c_n/\log(n)$ for all $i$.
Hoeffding's Inequality (Lemma~\ref{lem:hoeffding}) now shows
that for all $i < n$,
\beqan
    \Pr(z(i) \geq 1+\epsilon) &\leq& \exp\left(-2\epsilon^2/C(i,n)\right)  \\
        &\leq& \exp\left(-2\epsilon^2\log(n)/c_n\right).
\eeqan
Using the union bound,
\beqan
    \lefteqn{\limn \Pr\left( \max_{j=1,\ldots,n} z(i) > 1 + \epsilon\right) } \\
    &\leq&
       \limn  n\exp\left(- \frac{2\epsilon^2\log(n)}{c_n}\right) \\
     &=&  \limn  n^{1- 2\epsilon^2/c_n} = 0.
\eeqan
The final step is due to the fact that the technical condition
(\ref{eq:pjsqbnd}) in the theorem implies $c_n \arr 0$.  This proves the lemma.
\end{IEEEproof}

\subsection{Missed Detection Probability}
\label{sec:pmd}

Consider any $j \in \Itrue$.
Using (\ref{eq:residDef}) to rewrite (\ref{eq:rhotrue})
along with some algebra shows
\beqa
    \rhotrue(j) &=& \frac{|\abf_j'\Ptrue(j)\ybf|^2}{\|\Ptrue(j)\abf\|^2\|\Ptrue(j)\ybf_j\|^2}
        \nonumber \\
        &=& \frac{|\abf_j'(x_j\Ptrue(j)\abf_j + \ebf_j)|^2}
        {\|\Ptrue(j)\abf_j\|^2\|x_j\Ptrue(j)\abf_j+\ebf_j\|^2} \nonumber \\
           &\geq& \frac{s_j - 2\sqrt{z_js_j} + z_j}{s_j + 2\sqrt{z_js_j} + 1},
           \label{eq:rhotrueMD}
\eeqa
where
\beqa
    s_j &=& \frac{|x_j|^2\|\Ptrue(j)\abf_j\|^2}{\|\ebf_j\|^2}, \label{eq:sjdef} \\
    z_j &=& \frac{|\abf_j'\Ptrue(j)\ebf_j|^2}{\|\Ptrue(j)\abf_j\|^2\|\ebf_j\|^2}.
         \label{eq:zjdef}
\eeqa
Define
\[
    \smin = \min_{j \in \Itrue} s_j, \ \ \
    \zmax = \max_{j \in \Itrue} z_j.
\]
We will now bound $\smin$ from below and $\zmax$ from above.

We first start with $\smin$.   Conditional on $\xbf$ and $\Ptrue(j)$,
Lemma~\ref{lem:residDist} shows that each $\ebf_j$
is a spherically-symmetrically distributed Gaussian on the $m_j$-dimensional
range space of $\Ptrue(j)$.  Since there are asymptotically $\lambda n$ elements
in $\Itrue$, Lemma~\ref{lem:chiSqEq} along with (\ref{eq:lognm}) show that
\beq \label{eq:sjlima}
    \lim_{n \arr \infty} \max_{j \in \Itrue} \frac{m}{m_j\sigma^2(j)}\|\ebf_j\|^2 = 1,
\eeq
where the limit is in probability.
Similarly, $\Ptrue(j)\abf_j$ is also a spherically-symmetrically distributed Gaussian
in the range space of $\Ptrue(j)$.  Since $\Ptrue(j)$ is a projection from an $m$-dimensional
space to a $m_j$-dimensional space and $\Exp\|\abf_j\|^2 = 1$, we have that
$\Exp\|\Ptrue(j)\abf_j\|^2 = m_j/m$.
Therefore, Lemma~\ref{lem:chiSqEq} along with (\ref{eq:lognm}) show that
\beq \label{eq:sjlimb}
    \lim_{n \arr \infty} \min_{j \in \Itrue} \frac{m}{m_j}\|\Ptrue(j)\ebf_j\|^2 = 1.
\eeq
Taking the limit (in probability) of $\smin$,
\beqa
    \liminfn \frac{\smin}{\gamma} &=& \liminfn \min_{j \in \Itrue} \frac{s_j}{\gamma}
        \nonumber \\
    &\stackrel{(a)}{=}& \liminfn \min_{j \in \Itrue}
        \frac{|x_j|^2\|\Ptrue(j)\abf_j\|^2}{\gamma\|\ebf_j\|^2} \nonumber \\
    &\stackrel{(b)}{=}& \liminfn \min_{j \in \Itrue}
        \frac{|x_j|^2}{\gamma\sigma^2(j)} \nonumber \\
    &\stackrel{(c)}{=}& \liminfn \min_{j \in \Itrue}
        \frac{p_j}{\gamma\sigma^2(j)} \nonumber \\
    &\stackrel{(d)}{\geq}& \liminfn \min_{j \in \Itrue}
        \frac{p_j}{\gamma\sigmahat^2(j)} \nonumber \\
         &\stackrel{(e)}{\geq}& 1, \label{eq:sminLim}
\eeqa
where (a) follows from (\ref{eq:sjdef});
(b) follows from (\ref{eq:sjlima}) and (\ref{eq:sjlimb});
(c) follows from (\ref{eq:pjdef});
(d) follows from Lemma~\ref{lem:sigLim};
and (e) follows from (\ref{eq:sinrDef}).

We next consider $\zmax$.  Conditional on $\Ptrue(j)$, the vectors $\Ptrue(j)\abf_j$ and
$\ebf_j$ are independent spherically-symmetric Gaussians in the
range space of $\Ptrue(j)$.  It follows from Lemma~\ref{lem:betaProj} that
each $z_j$ is a $\BetaDist(1,m_j-1)$ random variable.
Since there are asymptotically
$\lambda n$ elements in $\Itrue$, Lemma~\ref{lem:betaLim} along
with (\ref{eq:miMin}) and (\ref{eq:lognm}) show that
\beq
    {\limsupn \frac{m - \lambda n}{2\log(\lambda n)} \zmax}
    =    \limsupn \frac{m - \lambda n}{2\log(\lambda n)} \max_{j \in \Itrue}
        z_j \leq 1. \label{eq:zmaxLim}
\eeq

The above analysis shows that for any $j \in \Itrue$,
\beqa
    \lefteqn{ \liminfn \min_{j \in \Itrue} \frac{1}{\sqrt{\mu}}(\sqrt{s_j} - \sqrt{z_j}) } \nonumber \\
    &\stackrel{(a)}{\geq}& \liminfn \frac{1}{\sqrt{\mu}}(\sqrt{\smin} - \sqrt{\zmax}) \nonumber \\
    &\stackrel{(b)}{\geq}& \liminfn \frac{1}{\sqrt{\mu}} \left(\sqrt{\gamma}
       - \sqrt{\frac{2\log(\lambda n)}{m-\lambda n}} \right) \nonumber  \\
    &\geq& \liminfn \sqrt{\frac{1+\delta}{\mu}}
        \left(\sqrt{\frac{\gamma}{1+\delta}}
       - \sqrt{\frac{2\log(\lambda n)}{m-\lambda n}} \right) \nonumber  \\
    &\stackrel{(c)}{\geq}& \liminfn \sqrt{\frac{2(1+\delta)}{(m-\lambda n)\mu}}
        \left(\sqrt{L(\lambda,n)}
       - \sqrt{\log(\lambda n)} \right) \nonumber  \\
    &\stackrel{(d)}{=}& \liminfn \sqrt{\frac{2(1+\delta)\log(n(1-\lambda))}{(m-\lambda n)\mu}}
        \nonumber  \\
    &\stackrel{(e)}{=}& \liminfn \sqrt{\frac{1+\delta}{1+\epsilon}} 
    \ \stackrel{(f)}{\geq} \ \sqrt{1+\epsilon} \label{eq:szdiff}
\eeqa
where (a) follows from the definitions of $\smin$ and $\zmax$;
(b) follows from (\ref{eq:sminLim}) and (\ref{eq:zmaxLim});
(c) follows from (\ref{eq:minMeasGam});
(d) follows from (\ref{eq:Ldef});
(e) follows from (\ref{eq:muDef}); and
(f) follows from (\ref{eq:epsDef}).
Therefore, starting with (\ref{eq:rhotrueMD}),
\beqan
    \lefteqn{
    \liminf_{n \arr \infty} \min_{j \in \Itrue}
    \frac{\rho(j)}{\mu} } \nonumber \\
    & \stackrel{(a)}{\geq}&
    \liminfn  \min_{j \in \Itrue} \frac{1}{\mu}
        \frac{s_j - 2\sqrt{z_js_j} + z_j}{s_j + 2\sqrt{z_js_j} + 1} \nonumber \\
    & =&
    \liminfn  \min_{j \in \Itrue} \frac{1}{\mu}
        \frac{(\sqrt{s_j} - \sqrt{z_j})^2}{s_j + 2\sqrt{z_js_j} + 1} \nonumber \\
    & \stackrel{(b)}{\geq}&
    \liminfn  \min_{j \in \Itrue}
        \frac{1+\epsilon}{s_j + 2\sqrt{z_js_j} + 1} \nonumber \\
    & \stackrel{(c)}{\geq}&
    \liminfn  \min_{j \in \Itrue}
        \frac{1+\epsilon}{s_j + 2\sqrt{s_j} + 1} \nonumber \\
    & \geq&
    \liminfn  \min_{j \in \Itrue}
        \frac{1+\epsilon}{\smin + 2\sqrt{\smin} + 1} \nonumber \\
    & \stackrel{(d)}{\geq}&
     \liminfn  \min_{j \in \Itrue}
        \frac{1+\epsilon}{(\sqrt{\gamma} + 1)^2}  \stackrel{(e)}{=}1+\epsilon,
\eeqan
where (a) follows from (\ref{eq:rhotrueMD});
(b) follows from (\ref{eq:szdiff});
(c) follows from the fact that $z_j \in [0,1]$ (it is a Beta distributed random variable);
(d) follows from (\ref{eq:sminLim}); and
(e) follows from the condition of the hypothesis of the theorem that $\gamma \arr 0$.
This proves the first requirement, condition (\ref{eq:rhoMD}).

\subsection{False Alarm Probability}
\label{sec:pfa}

Now consider any index $j \not \in \Itrue$.  This implies that $x_j=0$ and therefore
(\ref{eq:residDef}) shows that
\[
    \Ptrue(j)\ybf
     = \ebf_j.
\]
Hence from (\ref{eq:rhotrue}),
\beq \label{eq:rhozj}
   \rhotrue(j) = \frac{|\abf_j'\ebf|^2}
                      {\|\Ptrue(j)\abf\|^2\|\ebf_j\|^2}
   = z_j
\eeq
where $z_j$ is defined in (\ref{eq:zjdef}).  From the discussion above,
each $z_j$ has the $\BetaDist(2,m_j-2)$ distribution.
Since there are asymptotically $(1-\lambda) n$ elements in $\Itrue^c$,
the conditions (\ref{eq:miMin}) and (\ref{eq:lognm}) along with Lemma~\ref{lem:betaLim}
show that the limit
\beq \label{eq:zbndfa}
    \limsupn \max_{j \not\in \Itrue} \frac{m-\lambda n}{2\log(n(1-\lambda))}z_j \leq 1
\eeq
holds in probability.
Therefore,
\beqan
    \lefteqn{ \limsupn \max_{j \not\in \Itrue} \frac{1}{\mu}\rhotrue(j) } \\
    &\stackrel{(a)}{=}& \limsupn \max_{j \not\in \Itrue} \frac{1}{\mu}z_j  \\
    &\stackrel{(b)}{=}& \limsupn \max_{j \not\in \Itrue} \frac{m - \lambda n}{(1+\epsilon)
            \log(n(1-\lambda))}z_j \\
    &\stackrel{(c)}{\leq}& \frac{1}{1+\epsilon}
\eeqan
where (a) follows from (\ref{eq:rhozj});
(b) follows from (\ref{eq:muDef}); and
(c) follows from (\ref{eq:zbndfa}).
This proves (\ref{eq:rhoFA}) and thus completes the proof of the theorem.

\bibliographystyle{IEEEtran}
\bibliography{bibl}

\newcommand{\SortNoop}[1]{}
\begin{thebibliography}{10}
\providecommand{\url}[1]{#1}
\csname url@samestyle\endcsname
\providecommand{\newblock}{\relax}
\providecommand{\bibinfo}[2]{#2}
\providecommand{\BIBentrySTDinterwordspacing}{\spaceskip=0pt\relax}
\providecommand{\BIBentryALTinterwordstretchfactor}{4}
\providecommand{\BIBentryALTinterwordspacing}{\spaceskip=\fontdimen2\font plus
\BIBentryALTinterwordstretchfactor\fontdimen3\font minus
  \fontdimen4\font\relax}
\providecommand{\BIBforeignlanguage}[2]{{%
\expandafter\ifx\csname l@#1\endcsname\relax
\typeout{** WARNING: IEEEtran.bst: No hyphenation pattern has been}%
\typeout{** loaded for the language `#1'. Using the pattern for}%
\typeout{** the default language instead.}%
\else
\language=\csname l@#1\endcsname
\fi
#2}}
\providecommand{\BIBdecl}{\relax}
\BIBdecl

\bibitem{FletcherRG:09a}
A.~K. Fletcher, S.~Rangan, and V.~K. Goyal, ``A sparsity detection framework
  for on--off random access channels,'' in \emph{Proc. IEEE Int. Symp. Inform.
  Theory}, Seoul, Korea, Jun.--Jul. 2009, pp. 169--173.

\bibitem{Wainwright:09-ml}
M.~J. Wainwright, ``Information-theoretic limits on sparsity recovery in the
  high-dimensional and noisy setting,'' \emph{IEEE Trans. Inform. Theory},
  vol.~55, no.~12, pp. 5728--5741, Dec. 2009.

\bibitem{Wainwright:09-lasso}
------, ``Sharp thresholds for high-dimensional and noisy sparsity recovery
  using $\ell_1$-constrained quadratic programming (lasso),'' \emph{IEEE Trans.
  Inform. Theory}, vol.~55, no.~5, pp. 2183--2202, May 2009.

\bibitem{FletcherRG:09-IT}
A.~K. Fletcher, S.~Rangan, and V.~K. Goyal, ``Necessary and sufficient
  conditions for sparsity pattern recovery,'' \emph{IEEE Trans. Inform.
  Theory}, vol.~55, no.~12, pp. 5758--5772, Dec. 2009.

\bibitem{Natarajan:95}
B.~K. Natarajan, ``Sparse approximate solutions to linear systems,'' \emph{SIAM
  J. Computing}, vol.~24, no.~2, pp. 227--234, Apr. 1995.

\bibitem{MallatZ:93}
S.~G. Mallat and Z.~Zhang, ``Matching pursuits with time-frequency
  dictionaries,'' \emph{IEEE Trans. Signal Process.}, vol.~41, no.~12, pp.
  3397--3415, Dec. 1993.

\bibitem{ChenBL:89}
S.~Chen, S.~A. Billings, and W.~Luo, ``Orthogonal least squares methods and
  their application to non-linear system identification,'' \emph{Int. J.
  Control}, vol.~50, no.~5, pp. 1873--1896, Nov. 1989.

\bibitem{PatiRK:93}
Y.~C. Pati, R.~Rezaiifar, and P.~S. Krishnaprasad, ``Orthogonal matching
  pursuit: Recursive function approximation with applications to wavelet
  decomposition,'' in \emph{Conf.\ Rec.\ 27th Asilomar Conf.\ Sig., Sys., \&
  Comput.}, vol.~1, Pacific Grove, CA, Nov. 1993, pp. 40--44.

\bibitem{DavisMZ:94}
G.~Davis, S.~Mallat, and Z.~Zhang, ``Adaptive time-frequency decomposition,''
  \emph{Optical Eng.}, vol.~33, no.~7, pp. 2183--2191, Jul. 1994.

\bibitem{NeedellT:09}
D.~Needell and J.~A. Tropp, ``{CoSaMP}: Iterative signal recovery from
  incomplete and inaccurate samples,'' \emph{Appl. Comput. Harm. Anal.},
  vol.~26, no.~3, pp. 301--321, May 2009.

\bibitem{DaiM:09}
W.~Dai and O.~Milenkovic, ``Subspace pursuit for compressive sensing signal
  reconstruction,'' \emph{IEEE Trans. Inform. Theory}, vol.~55, no.~5, pp.
  2230--2249, May 2009.

\bibitem{ChenDS:99}
S.~S. Chen, D.~L. Donoho, and M.~A. Saunders, ``Atomic decomposition by basis
  pursuit,'' \emph{SIAM J. Sci. Comp.}, vol.~20, no.~1, pp. 33--61, 1999.

\bibitem{Tibshirani:96}
R.~Tibshirani, ``Regression shrinkage and selection via the lasso,'' \emph{J.
  Royal Stat. Soc., Ser. B}, vol.~58, no.~1, pp. 267--288, 1996.

\bibitem{CandesT:07}
E.~J. Cand{\`e}s and T.~Tao, ``The {D}antzig selector: Statistical estimation
  when $p$ is much larger than $n$,'' \emph{Ann. Stat.}, vol.~35, no.~6, pp.
  2313--2351, Dec. 2007.

\bibitem{DonohoET:06}
D.~L. Donoho, M.~Elad, and V.~N. Temlyakov, ``Stable recovery of sparse
  overcomplete representations in the presence of noise,'' \emph{IEEE Trans.
  Inform. Theory}, vol.~52, no.~1, pp. 6--18, Jan. 2006.

\bibitem{Tropp:04}
J.~A. Tropp, ``Greed is good: Algorithmic results for sparse approximation,''
  \emph{IEEE Trans. Inform. Theory}, vol.~50, no.~10, pp. 2231--2242, Oct.
  2004.

\bibitem{Tropp:06}
------, ``Just relax: Convex programming methods for identifying sparse signals
  in noise,'' \emph{IEEE Trans. Inform. Theory}, vol.~52, no.~3, pp.
  1030--1051, Mar. 2006.

\bibitem{CandesRT:06-IT}
E.~J. Cand{\`e}s, J.~Romberg, and T.~Tao, ``Robust uncertainty principles:
  Exact signal reconstruction from highly incomplete frequency information,''
  \emph{IEEE Trans. Inform. Theory}, vol.~52, no.~2, pp. 489--509, Feb. 2006.

\bibitem{Donoho:06}
D.~L. Donoho, ``Compressed sensing,'' \emph{IEEE Trans. Inform. Theory},
  vol.~52, no.~4, pp. 1289--1306, Apr. 2006.

\bibitem{CandesT:06}
E.~J. Cand{\`e}s and T.~Tao, ``Near-optimal signal recovery from random
  projections: Universal encoding strategies?'' \emph{IEEE Trans. Inform.
  Theory}, vol.~52, no.~12, pp. 5406--5425, Dec. 2006.

\bibitem{RanganFG:0x-IT}
S.~Rangan, A.~Fletcher, and V.~K. Goyal, ``Asymptotic analysis of {MAP}
  estimation via the replica method and applications to compressed sensing,''
  \emph{IEEE Trans. Inform. Theory}, 2011, to appear; available as
  arXiv:0906.3234v1 [cs.IT].

\bibitem{DonohoT:09}
D.~L. Donoho and J.~Tanner, ``Counting faces of randomly-projected polytopes
  when the projection radically lowers dimension,'' \emph{J. Amer. Math. Soc.},
  vol.~22, no.~1, pp. 1--53, Jan. 2009.

\bibitem{SarvothamBB:06-Allerton}
S.~Sarvotham, D.~Baron, and R.~G. Baraniuk, ``Measurements vs.\ bits:
  Compressed sensing meets information theory,'' in \emph{Proc. 44th Ann.
  Allerton Conf. on Commun., Control and Comp.}, Monticello, IL, Sep. 2006.

\bibitem{FletcherRG:07b}
A.~K. Fletcher, S.~Rangan, and V.~K. Goyal, ``Rate-distortion bounds for sparse
  approximation,'' in \emph{IEEE Statist. Sig. Process. Workshop}, Madison, WI,
  Aug. 2007, pp. 254--258.

\bibitem{Reeves:08}
G.~Reeves, ``Sparse signal sampling using noisy linear projections,'' Univ. of
  California, Berkeley, Dept.\ of Elec. Eng. and Comp. Sci., Tech. Rep.
  UCB/EECS-2008-3, Jan. 2008.

\bibitem{AkcakayaT:10}
M.~Ak{\c c}akaya and V.~Tarokh, ``Shannon-theoretic limits on noisy compressive
  sampling,'' \emph{IEEE Trans. Inform. Theory}, vol.~56, no.~1, pp. 492--504,
  Jan. 2010.

\bibitem{TroppG:07}
J.~A. Tropp and A.~C. Gilbert, ``Signal recovery from random measurements via
  orthogonal matching pursuit,'' \emph{IEEE Trans. Inform. Theory}, vol.~53,
  no.~12, pp. 4655--4666, Dec. 2007.

\bibitem{FletcherR:09}
A.~K. Fletcher and S.~Rangan, ``Orthogonal matching pursuit from noisy
  measurements: A new analysis,'' in \emph{Proc. Neural Information Process.
  Syst.}, Y.~Bengio, D.~Schuurmans, J.~Lafferty, C.~K.~I. Williams, and
  A.~Culotta, Eds., Vancouver, Canada, Dec. 2009.

\bibitem{RauhutSV:08}
H.~Rauhut, K.~Schnass, and P.~Vandergheynst, ``Compressed sensing and redundant
  dictionaries,'' \emph{IEEE Trans. Inform. Theory}, vol.~54, no.~5, pp.
  2210--2219, May 2008.

\bibitem{WangWR:10}
W.~Wang, M.~J. Wainwright, and K.~Ramchandran, ``Information-theoretic limits
  on sparse signal recovery: Dense versus sparse measurement matrices,''
  \emph{IEEE Trans. Inform. Theory}, vol.~56, no.~6, pp. 2967--2979, Jun. 2010.

\bibitem{DuarteSBWB:05}
M.~F. Duarte, S.~Sarvotham, D.~Baron, W.~B. Wakin, and R.~G. Baraniuk,
  ``Distributed compressed sensing of jointly sparse signals,'' in \emph{Conf.
  Rec. Asilomar Conf. on Signals, Syst. \& Computers}, Pacific Grove, CA,
  Oct.--Nov. 2005, pp. 1537--1541.

\bibitem{Tipping:01}
M.~Tipping, ``Sparse {B}ayesian learning and the relevance vector machine,''
  \emph{J. Machine Learning Research}, vol.~1, pp. 211--244, Sep. 2001.

\bibitem{WipfR:04}
D.~Wipf and B.~Rao, ``Sparse {B}ayesian learning for basis selection,''
  \emph{IEEE Trans. Signal Process.}, vol.~52, no.~8, pp. 2153--2164, Aug.
  2004.

\bibitem{SchniterPZ:0x}
P.~Schniter, L.~C. Potter, and J.~Ziniel, ``Fast {B}ayesian matching pursuit:
  Model uncertainty and parameter estimation for sparse linear models,''
  \emph{IEEE Trans. Signal Process.}, Aug. 2008, submitted.

\bibitem{WipfR:06}
D.~Wipf and B.~Rao, ``Comparing the effects of different weight distributions
  on finding sparse representations,'' in \emph{Proc. Neural Information
  Process. Syst.}, Vancouver, Canada, Dec. 2006.

\bibitem{AgrawalACM:05}
A.~Agrawal, J.~G. Andrews, J.~M. Cioffi, and T.~Meng, ``Iterative power control
  for imperfect successive interference cancellation,'' \emph{IEEE Trans.
  Wireless Comm.}, vol.~4, no.~3, pp. 878--884, May 2005.

\bibitem{EvansHP:00}
M.~Evans, N.~Hastings, and J.~B. Peacock, \emph{Statistical Distributions},
  3rd~ed.\hskip 1em plus 0.5em minus 0.4em\relax New York: John Wiley \& Sons,
  2000.

\bibitem{Hoeffding:63}
W.~Hoeffding, ``Probability inequalities for sums of bounded random
  variables,'' \emph{J. Amer. Stat. Assoc.}, vol.~58, no. 301, pp. 13--30, Mar.
  1963.

\end{thebibliography}

\end{document}